\documentstyle[prb,aps,multicol]{revtex}
  \renewcommand{\narrowtext}{\begin{multicols}{2} \global\columnwidth20.5pc}
  \renewcommand{\widetext}{\end{multicols} \global\columnwidth42.5pc}

 \input{psfig}

\begin{document}

\draft


\title{Nonlinearity of Acoustic Effects and High-Frequency
Electrical Conductivity in GaAs/AlGaAs Heterostructures under
Conditions of the Integer Quantum Hall Effect}

\author{I. L. Drichko$^a$, A. M. Diakonov$^a$, I. Yu. Smirnov$^a$,
 and A. I. Toropov$^b$ }
\address{$^a$A. F. Ioffe Physicotechnical Institute of Russian
Academy of Sciences, Polytekhnicheskaya 26, 194021, St.Petersburg,
Russia;\\ $^b$Semiconductors Physics Institute of Siberian
Division of Russian Academy of Sciences, Ak. Lavrentieva 13,
630090,  Novosibirsk, Russia}

\date{\today} \maketitle

\begin{abstract}
The absorption coefficient for surface acoustic wave $\Gamma$ and
variation in the wave velocity $\Delta V/V$ were measured in
GaAs/AlGaAs heterostructures; the above quantities are related to
interaction of the wave with two-dimensional electron gas and
depend nonlinearly on the power of the wave. Measurements were
performed under conditions of the integer quantum Hall effect
(IQHE), in which case the two-dimensional electron gas was
localized in a random fluctuation potential of impurities. The
dependences of the components $\sigma_1(E)$ and $\sigma_2(E)$ of
high-frequency conductivity $\sigma=\sigma_1 - i\sigma_2$ on the
electric field of the surface wave were determined. In the range
of the conductivity obeying the Arrhenius law ($\sigma_1 \gg
\sigma_2$), the results obtained are interpreted in terms of the
Shklovskii theory of nonlinear percolation-based conductivity,
which makes it possible to estimate the magnitude of the
fluctuation potential of impurities. The dependences $\sigma_1(E)$
and $\sigma_2(E)$ in the range of high-frequency hopping
electrical conductivity, in which case ($\sigma_1 \ll \sigma_2$)
and the theory of nonlinearities has not been yet developed, are
reported.
\end{abstract}
\narrowtext

\section{Introduction}

Studies of the kinetic effects in GaAs/AlGaAs heterostructures
with two-dimensional (2D) electron gas in strong constant electric
fields \cite{1,2,3,4} show that nonlinear effects are adequately
accounted for by heating of electron gas. The main issue of
controversy in the interpretation of the above results is related
to identification of the relevant mechanism of electron-energy
relaxation. The theory of electron-gas heating and the
energy-relaxation mechanisms in the 2D case were considered in
Ref.~\onlinecite{5}. In Ref.~\onlinecite{6}, the dependence of the
coefficient of absorption of a surface acoustic wave (SAW) by 2D
electron gas in a GaAs/AlGaAs heterostructure on the SAW intensity
was also explained by heating of 2D electron gas by the SAW
alternating electric field; it was also shown that the
electron-energy relaxation time is controlled by energy
dissipation at the piezoelectric potential of acoustic phonons
under conditions of strong screening.

In the magnetic-field range where electrons are localized (i.e..
under the conditions of IQHE), kinetic effects were also actively
studied in a strong constant electric field \cite{7,8,9,10,11}.
However, notwithstanding the fact that nonlinear dependences of
current on voltage were similar in all cases, these dependences
have not been unambiguously interpreted so far. Thus, these
nonlinearities were explained by the heating of 2D electron gas
\cite{7,8,9}. by impurity breakdown in homogeneous
 electric field \cite{8}, and by resonance tunneling of electrons
 between the Landau levels \cite{10};
 still another explanation was based on the theory of
 variable-range hopping conduction in a strong electric field \cite{11,12}.
 In Ref.~\onlinecite{13}, in order to explain the nonlinearities in the
dependence of the width of the IQHE step on the current density in
a GaAs/AlGaAs heterostructure at $T$ = 2.05 K. the model of
heating of 2D electron gas was used. It was found that the
dependence of electron temperature $T_e$ on the current $I$
coincided with $T_e (I)$ measured in the absence of a magnetic
field. As a plausible reason for this behavior, the authors of
Ref.~\onlinecite{13} could only suggest that it arose from the
injection of hot charge carriers from near-contact regions where
the Hall voltage was shorted out by the current contacts.

In connection with the above, the acoustic method seems to hold
considerable promise for studying the nonlinear effects under the
conditions of IQHE; this method is particularly convenient owing
to the fact that there is no need for electrical contacts in such
measurements.

In this work, we studied the dependences of absorptivity $\Gamma$
and variations $\Delta V/V$ in the velocity of SAWs in a
piezoelectric due to the interaction of SAWs with 2D electron gas
in GaAs/AlGaAs heterostructures on the SAW power $W$ absorbed in
the sample (or on the SAW electric field $E$) under the conditions
of IQHE ($T$ = 1.5 K), which corresponds to the
carrier-localization domain. The experimental dependences
$\Gamma$(E) and $\Delta V/V (E)$ were used to calculate the real
and imaginary components of high-frequency electrical conductivity
$\sigma_1(E)$ and $\sigma_2(E)$), as the high-frequency
conductivity is written in the complex form \cite{14} $\sigma(E) =
\sigma_1(E) - i \sigma_2(E)$ under the electron-localization
conditions. The mechanisms of the nonlinearities were studied by
analyzing the dependences of the components of the high-frequency
electrical conductivity on the strength (i.e., on absorbed power)
of a high-frequency electric field.

\section{EXPERIMENTAL METHODS AND RESULTS} \label{es}

In this work, the effect of the SAW power introduced into sample
($f$= 30 MHz) on the absorption coefficient $\Gamma$ and on the
relative variation in the SAW velocity $\Delta V/V$ was measured
at $T$=1.5K in magnetic fields corresponding to the midpoint of
the Hall plateau (i.e., under the conditions of IQHE). We used the
GaAs/AlGaAs heterostructures $\delta$-doped with silicon; the
density of electrons in 2D gas in the channel was $n = (1.3-2.7)
\times 10^{11}cm^{-2}$ and their mobility was $\mu \approx 2
\times 10^5 cm^2/(V\cdot s)$. The heterostructures were grown by
molecular-beam epitaxy and involved a spacer layer $4\times
10^{-6}$cm in width. The experimental procedure was described in
detail elsewhere \cite{15}. Here, we only note
 that the structure with 2D electron gas was grown on the piezoelectric
 (lithium niobate) surface over which the SAWs propagated.
 An alternating electric field having the frequency of SAW and
 accompanying the strain wave penetrates into the channel with 2D electron
 gas, induces electric currents and, correspondingly, introduces ohmic
 losses. As a result of such an interaction, the wave energy is absorbed.
 Experimentally, we measured the absorption coefficient $\Gamma$ and
 the relative variation in the SAW velocity $\Delta V/V$ in relation to the
 magnetic induction. Since the measured quantities $\Gamma$ and $\Delta V/V$
 are defined by the electric conductivity of 2D electron gas,
 electron-spectrum quantization resulting in the Shubnikov-de Haas
 oscillations brings about oscillations in the above effects as well.

Figure 1 shows the dependences of $\Gamma$ and $\Delta V/V$ on the
power $P$ at the output of the RF generator (for a frequency of
$f$= 30 MHz) for the filling numbers $\nu$ = 2; 4 and 6 for a
sample with the 2D-electron gas density of $n = 2.7 \times 10^{11}
cm^{-2}$ Here, $\nu = nch/eH$, where $H$ is the magnetic field
strength. In the insert in Fig. 1, the dependences of $\Gamma$ and
$\Delta V/V$ on the magnetic field are shown for several values of
the SAW power at a temperature of 1.5K. The form of the
dependences of $\Gamma$ on the magnetic field is analyzed in
detail in Ref.~\onlinecite{15}: positions of the peaks in the
$\Gamma$(H) and $\Delta V/V(H)$ curves are equidistant in $1/H$,
and the splitting of the $\Gamma (Í)$ peaks under the IQHE
conditions is related to its relaxation origin. It can be seen
from Fig. 1 that an increase in the supplied SAW power results
invariably in a decrease in $\Delta V/V$ irrespective of the
filling number, which corresponds to an increase in electrical
conductivity \cite{6}. The form of dependences $\Gamma(H)$ for
dissimilar filling numbers is different: absorption increases with
increasing $P$ for $\nu$ = 2, whereas, for $\nu$ = 4 and 6,
$\Gamma$ increases initially as the SAW power increases, attains a
maximum, and then decreases. The smaller the filling number, the
higher the power corresponding to the maximum in $\Gamma$. It is
also evident from Fig. 1 that the higher the magnetic field (the
smaller the filling number), the higher is power $P$ corresponding
to the onset of the dependences $\Gamma$(P) and $\Delta V/V(P)$.

\begin{figure}[h]
\centerline{\psfig{figure=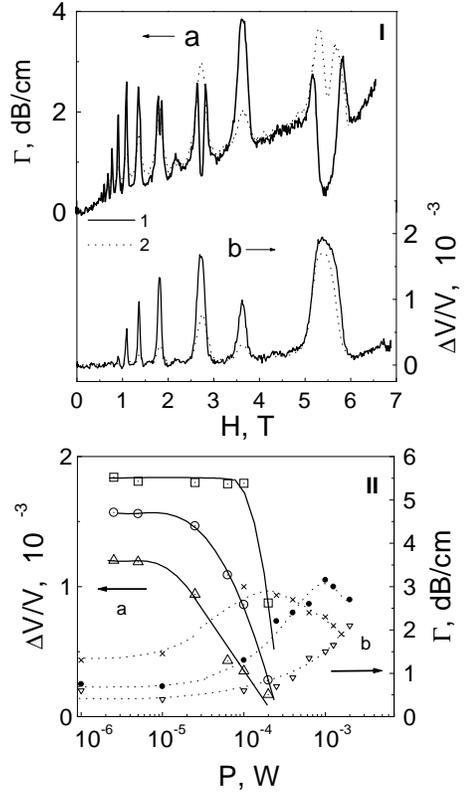,width=6cm,clip=}}
\caption{I - Dependences of (a) $\Gamma$ and
(b) $\Delta V/V$ on the magnetic field $H$ at $T$= 1.5 K for power $P$
at the generator output equal to (a) $10^{-5}$ and (b) $2 \times 10^{-6}$ W
(shown by solid lines) and equal to (a) $2 \times 10^{-3}$ and
(b) $10^{-4}$ W (shown by dashed lines).
II - Dependences of (a) relative velocity variation $\Delta V/V$ and
(b) absorption coefficient $\Gamma$ of a surface acoustic wave on
power $P$ at the output of a generator in magnetic fields of
($\Box$ and $\bigtriangledown$) 5.5, ($\circ$ and $\bullet$) 2.7, and
($\triangle$ and $\times$) 1.8 T.
 \label{f_1}}
\end{figure}

\section{DISCUSSION OF THE RESULTS}\label{er}

In the experimental configuration of this work, the quantities
$\Gamma$ and $\Delta V/V$ are defined \cite{16,17} by the
formulas
\begin{eqnarray}
\Gamma=8.68 \frac{K^2}{2} qA\frac { (\frac{4\pi \sigma_{1}}
{\varepsilon_sV})t(q) } {[1+(\frac{4\pi \sigma_{2}}{\varepsilon_s
V})t(q)]^2+ [(\frac{4\pi \sigma_{1}}{\varepsilon_s V})t(q)]^2},
\frac{dB}{cm} \label{eq1}
\end{eqnarray}
$$A=8b(q)(\varepsilon_1+\varepsilon_0)\varepsilon_0^2\varepsilon_s
e^{(-2q(a+d))},$$ $$ \frac{\Delta V}{V}= \frac{K^2}{2}
A\frac{(\frac{4\pi \sigma_{2}}{\varepsilon_s V})t(q)+1}
{[1+(\frac{4\pi \sigma_{2}}{\varepsilon_s V})t(q)]^2+ [(\frac{4\pi
\sigma_{1}}{\varepsilon_s V})t(q)]^2}, $$
$$b(q)=[b_1(q)[b_2(q)-b_3(q)]]^{-1,}$$
$$t(q)=[b_2(q)-b_3(q)]/[2b_1(q)],$$
$$b_1(q)=(\varepsilon_1+\varepsilon_0)(\varepsilon_s+\varepsilon_0)
-(\varepsilon_1-\varepsilon_0)
(\varepsilon_s-\varepsilon_0)e^{(-2qa)},$$
$$b_2(q)=(\varepsilon_1+\varepsilon_0)(\varepsilon_s+\varepsilon_0)+
(\varepsilon_1+\varepsilon_0)(\varepsilon_s-\varepsilon_0)e^{(-2qd)},$$
\begin{eqnarray}
b_3(q)=(\varepsilon_1-\varepsilon_0)(\varepsilon_s-\varepsilon_0)e^{(-2qa)}
+(\varepsilon_1-\varepsilon_0)(\varepsilon_s+\varepsilon_0)\times
\nonumber & \\ \times e^{[-2q(a+d)]},\nonumber
\end{eqnarray}

where the absorption coefficient $\Gamma$ is expressed in dB/cm;
$K^2$ is the piezoelectric constant of $LiNbO_3$; $q$ and $V$ are
the wave vector and the velocity of SAW, respectively; $a$ is the
distance between the dielectric and the heterostructure under
consideration; $d$ is the depth of the position of the 2D
electron-gas layer; $\varepsilon_1$, $\varepsilon_0$ and
$\varepsilon_s$ are the permittivities of lithium niobate, vacuum,
and gallium arsenide, respectively; and $\sigma_1$ and $\sigma_2$
are the components of the complex high-frequency electri-cal
conductivity of 2D electron gas: $\sigma=\sigma_1 - i\sigma_2$.
The necessity of considering both components of high-frequency
conductivity was demonstrated in \cite{14} and was related to
localization of electrons under the conditions of IQHE. These
formulas make it possible to determine $\sigma_1$ and $\sigma_2$
from the quantities $\Gamma$ and $\Delta V/V$ measured
experimentally.


\begin{figure}[h]
\centerline{\psfig{figure=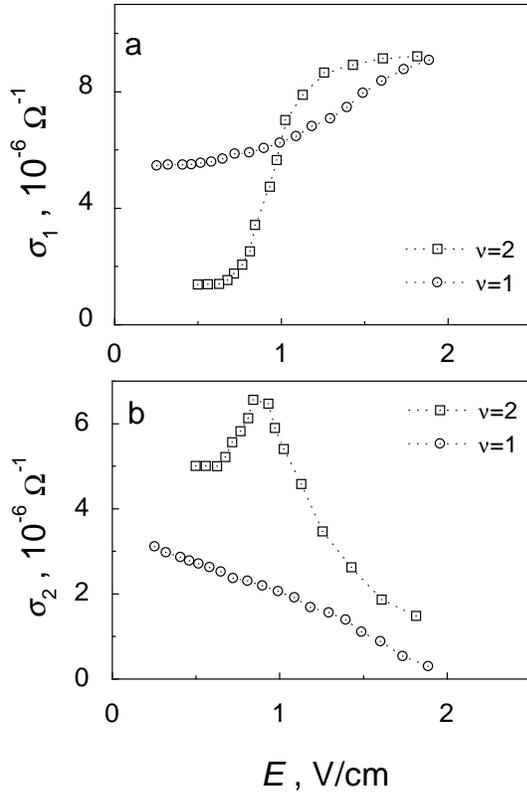,width=7cm,clip=} }
\caption{Dependences of (a) real $\sigma_1$ and (b) imaginary $\sigma_2$
components of high-frequency electrical conductivity on electric field $E$
for sample AG106 ($n = 1.3 \times 10^{11} cm^{-2}$) at $T$=1.5K.
\label{s12E}}
\end{figure}

Figure 2 shows the dependences of $\sigma_1$ and $\sigma_2$ on the
strength of high-frequency electric field $E$ accompanying the SAW
for the sample with $n = 1.3 \times 10^{11} cm^{-2}$ and the
filling numbers $\nu$ = 2 and 1. It can be seen from Fig. 2 that,
for $\nu$ = 2 (orbital splitting), $\sigma_1$ and $\sigma_2$
initially increase with increasing electric field and, beginning
with a certain $E$, $\sigma_2$ decreases rapidly while $\sigma_1$
 continues to increase. In magnetic fields
corresponding to the spin splitting ($\nu$ = 1), $\sigma_1$
increases and $\sigma_2$ decreases with increasing $E$.

Figure 3 shows the dependences of $\sigma_1$ and $\sigma_2$ on the
strength of high-frequency electric field E for the sample with $n
= 2.7 \times 10^{11} cm^{-2}$ for the filling numbers of $\nu$= 2;
4 and 6. It can be seen from Fig. 3 that, in the magnetic field
$H$ amounting to 5.5T ($\nu$=2), both components of high-frequency
conductivity are independent of $E$ in a wide range of electric
fields; for fields higher than a certain value $E_1$ these
components increase with $E$. In a magnetic field of $H$ = 2.7 T
($\nu$= 4), an increase in $\sigma_1$ and $\sigma_2$ sets in an
electric field $E_2 < E_1$, and, for $H$ = 1.8T ($\nu$ = 6),
$\sigma_2$ increases initially with increasing $E$, attains a
maximum, and then decreases (similarly to what is shown in Fig. 2
for $H$= 2.7 T in the case of $\nu$ = 2).

\input{f_3.inp}

The high-frequency electric field of SAW is calculated here with
the formula reported in Ref.~\onlinecite{6}. The only difference
is that we have $\sigma= \sigma_1-i\sigma_2$ i.e.,

\begin{eqnarray}
\  |E|^2=K^2\frac{32\pi}{V}(\varepsilon_1+\varepsilon_0)
\frac{zqe^{(-2q(a+d))}} {(1+\frac{4\pi \sigma_{2}}{\varepsilon_s
V}t)^2+(\frac{4\pi \sigma_{1}}{\varepsilon_s V}t)^2}W,\label{eq2}
\end{eqnarray}

\begin{eqnarray}
z=[(\varepsilon_1 + \varepsilon_0)(\varepsilon_s + \varepsilon_0)-
e^{(-2qa)}(\varepsilon_1-\varepsilon_0) \nonumber & \\ \times
(\varepsilon_s-\varepsilon_0)]^{-2}, \nonumber
\end{eqnarray}

In order to explain the above dependences of $\sigma_1$ and
$\sigma_2$ on E, we should rely on the fact that, as was shown in
Ref.~\onlinecite{14}, electrical conduction in the temperature
range of $T$= 1.5-4.2 K is simultaneously governed by the
following two mechanisms: (i) the Arrhenius-type mechanism related
to activation of charge carriers from the Fermi level, where these
carriers are in localized states, to the percolation level and
(ii) the mechanism of hop-ping over localized states in the
vicinity of the Fermi level. For $T$ = 1.5 K, the contributions of
these two mechanisms vary in relation to the filling number (the
magnetic field). The smaller the filling number (the higher the
magnetic field), the larger the activation energy defined by the
value of $0.5\hbar\omega_c$ ($\omega_c$ is the cyclotron
frequency) and the smaller the Arrhenius contribution to the
conductivity. In the case of hopping high-frequency conduction,
the imaginary component has the value of $\sigma_2 \approx 10
\sigma_1 \gg \sigma_1$ \cite{14} and begins to decrease with an
increasing number of delocalized electrons as a result of the
activation process. Thus, the ratio $\sigma_1 / \sigma_1$ is a
quantity characterizing the contribution of the above two
mechanisms to conduction: if $\sigma_1 / \sigma_1 \approx 1$, the
hopping conduction is dominant, whereas, if $\sigma_1 / \sigma_1
\gg 1$, the Arrhenius-law conduction is dominant and the hop-ping
conduction may be ignored.

In view of the above, the nonlinearities should be analyzed
separately for two different domains.

\subsection{Nonlinearities in the Region of Arrhenius-Type
Conduction ($\sigma_1 / \sigma_2 \gg 1$)}\label{cc}

The influence of a strong constant electric field on electrical
conductivity stemming from activation of charge carriers to the
percolation level of the conduction band distorted by random
fluctuation potential of charged impurities was considered by
Shklovskii in Ref.~\onlinecite{18}. In fact, we study here the
influence of a strong electric field on conduction over the
percolation level, with the role of the electric field limited to
a reduction of activation energy, which may be interpreted as a
lowering of the percolation threshold.

In this case, an increase in electrical conductivity in a strong
electric field with decreasing activation energy is given by

\begin{equation}\label{eq3}
  \sigma_1 / \sigma_1^0 = exp[(CeElV_0)^{1/(1+\gamma)}/kT],
\end{equation}

where $\sigma_1^0$ is the conductivity in the linear mode, $E$ is
the electric field strength, $T$ is temperature, $Ñ$ is a
numerical coefficient, $V_0$ is the amplitude of fluctuations in
the potential-relief pattern (the characteristic spatial scale of
a potential), and $\gamma$ is the coefficient that depends on
dimensionality: $\gamma$= 0.9 for the three-dimensional (3D) case
and $\gamma$= 4/3 for the 2D case \cite{19}.

Thus, in the 2D case under consideration, formula \ref{eq3} can be
rewritten as

\begin{equation}\label{eq4}
  \sigma_1 / \sigma_1^0 = exp(\alpha E^{3/7}/kT),
\end{equation}

where

\begin{equation}\label{eq5}
\alpha=(Cel_{sp} V_0)^{3/7},
\end{equation}

and $l_{sp}$ is the characteristic spatial scale of the potential,
which may be taken as equal to the spacer thickness \cite{20} in
the heterostructures we studied ($l_{sp} = 4 \times 10^{-6} cm$).

In the experiment we performed, the following conditions were
satisfied:

\begin{equation}\label{eq6}
ql_{sp} \ll 1, \omega\tau \ll 1.
\end{equation}

Here, $q$ and $\omega$ are the wave vector and frequency of SAW,
respectively, and $\tau$ is the electron-momentum relaxation time.
Therefore, we may regard the wave as standing and we can use the
formulas obtained for the constant electric field in the analysis
of dependences of high-frequency conductivity on the alternating
SAW electric field.

Figure 4 shows the dependences of $\sigma_1 / \sigma_2$ on
electric-field strength $E$ calculated according to \ref{eq2} for
two samples in different magnetic fields. It can be seen from Fig.
4 that the condition $\sigma_1 / \sigma_2$ is met for sample AG106
alone; therefore, the dependences $\sigma_1 / \sigma_1^0$ on $E$
can be analyzed on the basis of \cite{18} only for this sample.
Figure 5 shows the dependence of $\ln \sigma_1 / \sigma_1^0$ on
$E^{3/7}$. It can be seen that the dependence is linear; the slope
of the corresponding straight line makes it possible to determine
(to within a numerical factor) $V_0$ i.e., the amplitude of
fluctuations in the potential relief pattern. We found that
$V_0\approx 1.5 meV$.

Calculation of the fluctuation amplitude by the formula \cite{21}

\begin{equation}\label{eq7}
  V_0=(e^2/\varepsilon_s)/\sqrt{n},
\end{equation}

where $n$ is the density of ionized impurities equal in our case
to the carrier density in the 2D channel with $n = 1.3\times
10^{11} cm^{-2}$ yields $V_0$ = 4.5 meV, which coincides by an
order of magnitude with $V_0$ determined experimentally to within
a numerical factor.

\input{f_4.inp}

It is stated in Ref.~\onlinecite{18} that the electric-field range
in which formula \ref{eq3} is valid is limited by the inequalities

\begin{equation}\label{eq8}
  V_0 \gg eEl_{sp} \gg kT(kT/V_0)^{4/3}.
\end{equation}

Using the experimental value of $V_0$, we can estimate the
quantities in this inequality:

$1.5 meV \gg 3\times 10^{-3} meV \approx 5 \times 10^{-3} meV$ at
the threshold of nonlinearities; and

$1.5 meV \gg 7 \times 10^{-3} meV > 5 \times 10^{-3} meV$ at the
upper limit in our measurements.

Taking into account that the fluctuation amplitude is determined
to within a numerical factor, we may consider that the inequality
holds and we can use the theory \cite{18} to interpret
nonlinearities arising in the case where the conductivity obeys
the Arrhenius law.

\input{f_5.inp}

It is worth noting here that the nonlinearities under a constant
electric field $E$ were studied for $E >$ 5-10 V/cm.

The results reported here were obtained for electric fields lower
than 2 V/cm. Thus, the following general pattern emerges. In
electric fields $E \approx$ 1 V/cm corresponding to the
prebreakdown region, electrical conductivity increases owing to an
increase in concentration in the empty Landau band due to
activation of localized electrons from the Fermi level; in this
case, the activation energy depends on the electric field (the
higher the electric field, the lower the activation energy). In
electric fields $E \approx$ 10 V/cm, a sharp increase in electron
concentration in the empty Landau band due to the impurity
breakdown sets in (this phenomenon was clearly observed, for
example, in Ref.~\onlinecite{9}). Electrons brought into
delocalized states of the Landau band are heated in a strong
electric field applied to the sample.

\subsection{Nonlinearities in the Region of
High-Frequency Hopping Conduction  ($\sigma_1 / \sigma_2 \approx
0.1$)}\label{cd}

In the case of electrical conductivity limited by activation, we
could use the nonlinearity theory developed for constant fields to
interpret the nonlinearities in high-frequency conductivity,
because this theory described the influence of a strong electric
field on the motion of quasi-free electrons activated to the
percolation level. The mechanism of conduction in a constant
electric field and that in a high-frequency field are found to be
the same. However, in the case where the electrons are localized,
the mechanisms of high-frequency hopping conduction and dc hopping
conduction are different: in the dc mode, the conduction is
accomplished by hops of electrons between two edges of the sample,
whereas, in a high-frequency electric field with electrons
localized at separate impurity atoms, conduction can be effected
by hops of electrons between two impurity atoms separated by a
distance smaller than the average one (within an impurity pair
with a single electron); in this case, transitions of electrons
between different pairs do not occur (a neighbor-site model).
Therefore, it is understandable that the theory developed for
nonlinearities in the mode of dc hopping conduction \cite{12}
cannot be used to interpret the nonlinearities we observed in
high-frequency hopping conduction (Figs. 2,3).

The dependences of resistivities $\rho_{xx}$ and $\rho_{xy}$, on
the current density in a GaAs/AlGaAs heterostructure were observed
under the IQHE conditions in the region of variable-range hopping
conduction at $T<$1K \cite{11}. These dependences were analyzed
using the theory \cite{12} for variable-range hopping conduction
in a strong electric field for 2D electron gas under conditions of
the quantum Hall effect. An introduction of effective temperature
for hopping conduction in terms of Ref.~\onlinecite{12} makes it
possible to use the above measurements to determine the value of
localization length. However, it was found that the value thus
obtained was by almost an order of magnitude larger than the
localization length deter-mined from the temperature dependence of
$\rho_{xx}$ in a linear conduction mode. This fact was related
\cite{11} to non-uniformity in the distribution of the electric
field.

The theory of nonlinear high-frequency electrical conduction was
developed in Ref.~\onlinecite{22} for the 3D case. However, the
dependences $\Delta\sigma_{1,2}^{nl} / \sigma_{1,2}^0 \propto W^2$
we observed are stronger than those predicted in
Ref.~\onlinecite{21} ($\Delta \sigma \propto W$). Here,
$\sigma_{1,2}^0$ is the conductivity in linear mode,
$\Delta\sigma_{1,2}^{nl}=\sigma_{1,2}(E)-\sigma_{1,2}^0$,
$\sigma_{1,2}(E)$ is the conductivity measured experimentally, and
$W$ is the SAW power absorbed in the sample. At present, the
absence of a theory for the 2D case prevents us from analyzing the
obtained results. The dependences of or,$\sigma_1$ and $\sigma_2$
on the SAW electric field under conditions of hopping conduction
set in under the electric field, which becomes lower as the
magnetic-field increases; this fact is qualitatively well
explained by assuming that the magnetic field affects the overlap
integral for localized states at different impurities and, thus,
brings about a depression of hopping conduction and a decrease in
the relative nonlinear component \cite{22}.

Dependence of absorpitivity of SAW by 2D electron gas on the SAW
power in GaAs/AlGaAs heterostructures was previously observed
\cite{23,24} in magnetic fields corresponding to small integer
filling numbers when the Fermi level was in midposition between
the neighboring Landau levels and when the electrons were
localized. However, it is hardly possible to accept the
interpretation \cite{23,24} of these dependences as due to the
heating of 2D electron gas, which, as was mentioned above,
manifests itself only in the case where the electrons in 2D
configuration are delocalized.

\section{Conclusion}\label{cf}

We studied nonlinear dependences of absorptivity and variation in
the SAW velocity induced by 2D electron gas on the intensity of
sound under the conditions of the integer quantum Hall effect in
GaAs/AlGaAs heterostructures.

The nonlinearities were analyzed on the basis of high-frequency
electrical conductivity that had a complex form and was calculated
from experimental data.

It is shown that the electric-field range corresponding to
nonlinearities can be divided into two domains.

In the electric-field domain where $Re \sigma = \sigma_1 \gg Im
\sigma =\sigma_2$, The dependence $ \sigma_1 (E)$ is not only
adequately accounted for by the Shklovskii dc nonlinearity theory
\cite{18} describing the influence of a strong electric field on
the motion of quasi-free electrons activated to the percolation
level, but also makes it possible to evaluate the magnitude of the
fluctuation impurity potential.

In the electric-field domain where $Im \sigma =\sigma_2 \gg Re
\sigma = \sigma_1$. the 2D high-frequency hopping conduction
apparently takes place; as of yet, the nonlinearity theory for
this type of conduction has not been developed.

\section{Acknowledgments}

We thank V.D. Kagan for his helpful participation in discussions
of the results.

This work was supported by the Russian Foundation for Basic
Research (project No. 98-02-18280) and the Ministry of Science
(project No. 97-1043).

\widetext
 \end{document}